Commentary
**The *in silico* macrophage: toward a better understanding of inflammatory disease**

Peter Ghazal[1], Steven Watterson[1], Kevin Robertson[1] and David C Kluth[2]

[1]Division of Pathway Medicine and Centre for Systems Biology Edinburgh, University of Edinburgh, Chancellor's Building, Little France Crescent, Edinburgh EH16 4SB, UK
[2]MRC Centre for Inflammation Research, University of Edinburgh, The Queen's Medical Research Institute, Little France Crescent, Edinburgh EH16 4TJ, UK

*Correspondence: P.ghazal@ed.ac.uk

**Abstract**
Macrophages function as sentinel, cell-regulatory 'hubs' capable of initiating, perpetuating and contributing to the resolution of an inflammatory response, following their activation from a resting state. Highly complex and varied gene expression programs within the macrophage enable such functional diversity. To investigate how programs of gene expression relate to the phenotypic attributes of the macrophage, the development of *in silico* modeling methods is needed. Such models need to cover multiple scales, from molecular pathways in cell-autonomous immunity and intercellular communication pathways in tissue inflammation to whole organism response pathways in systemic disease. Here, we highlight the potential of *in silico* macrophage modeling as an amenable and important yet under-exploited tool in aiding in our understanding of the immune inflammatory response. We also discuss how *in silico* macrophage modeling can help in future therapeutic strategies for modulating both the acute protective effects of inflammation (such as host defense and tissue repair) and the harmful chronic effects (such as autoimmune diseases).

**The complex macrophage and its role in inflammation**

> *'We can only see a short distance ahead, but we can see plenty there that needs to be done.'*
> Alan Turing

Macrophages are a key cell type involved in all stages of the inflammatory response and have diverse functions that are important in the response to injury or infection and the resolution of inflammation. Significantly, under certain circumstances, macrophages can also propagate injury. Macrophages have a front-line role in host defense, where they respond directly to microbes and host danger signals far more sensitively than non-professional innate immune cells. As professional antigen presenting cells they are also an important link between the activation and the coordination of the adaptive immune system.

The response of a macrophage to tissue damage or pathogen insult is mediated by pattern recognition receptors that trigger pathways leading to the production of pro-inflammatory cytokines and chemokines. Typically, this happens either through Toll-like receptor pathways, leading to production of tumor necrosis factor (TNF)-$\alpha$, interleukin (IL)-6 and IL-12, or inflammasome-activation pathways, leading to production of IL-1$\beta$ and IL-18. These and other associated outputs orchestrate the influx of further inflammatory cells and reactive oxygen and nitrogen species that promote microbial killing and phagocytosis.

Notably, activated macrophages can mediate inflammatory diseases during, for example, an acute infection (e.g. sepsis) or via chronic autoimmune inflammation (such as atherosclerosis, rheumatoid arthritis or glomerulonephritis). Inappropriate activation under these circumstances can only be understood by investigating how temporal intracellular and intercellular molecular behavior relates to local and systemic. Sepsis remains an important medical problem, with high mortality and still no effective targeted therapies [1], so there is a great need for new, systems-level insights into how normally protective immune responses develop into life-threatening diseases. Atherosclerosis has similar systems-level traits, as a wide variety of immune and metabolic system components are associated with the pathogenesis of plaques [2].

Macrophages infiltrate tissues following most forms of injury, including infection, ischemia and trauma, and most types of autoimmune inflammation, such as rheumatoid arthritis, glomerulonephritis, diabetes mellitus, cardiovascular disease and multiple sclerosis (for a recent review, see [3]). Hence, they are key participants in nearly all forms of inflammatory injury and often function as regulatory hubs under these circumstances. They are, therefore, an attractive choice for *in silico* modeling and the development of computational methods targeted at enhancing inflammation research.

As a regulatory hub, macrophages have important roles in the regulation and resolution of inflammation. In skin wounds, switching the phenotype of infiltrating macrophages to express heme oxygenase 1 is associated with the resolution of inflammation, and inhibition of heme oxygenase 1 delays wound healing. Resolution of inflammation also requires removal of apoptotic cells and other debris. Macrophage uptake of apoptotic cells, including neutrophils, results in an anti-inflammatory phenotype marked by production of IL-10, transforming growth factor (TGF)-$\beta$ and anti-inflammatory lipoxins [4]. All of these functions provide therapeutically relevant starting points for developing predictive multi-scale models.

Here, we highlight the potential of *in silico* modeling of macrophage biology as an important tool for understanding the complexity of inflammatory injury and repair.

**The *in silico* macrophage**
*In silico* analysis brings together comprehensive data from the literature and high-throughput studies. It can produce not only static pictures of the interactions along pathways, but, following conversion to a dynamic mathematical model can also, more powerfully, yield a predictive picture of the mechanisms controlling phenotype. However, the caveat to this is that a model can only be as good as the data that are used to build it.

Statistical analyses of genome-wide transcriptional activity and protein abundance have been used to infer the critical genes and pathways involved in the response to specific cellular challenges (for example, [5,6]). Significantly, however, such research synthesis efforts collectively produce static relationship-based graphs with only tentative descriptions of pathway function. In isolation, such graphs cannot distinguish between causal and correlated activity. Macrophage resources have also been compiled and are presented in several websites [7-9]. As a consequence of these efforts, significant portions of the cellular pathways of macrophages have been compiled from the published literature and presented as maps [10-12]. It is recognized, however, that ensuring coherence of cell types, treatments, experimental methods and statistically robust data across the primary studies is a challenge.

Developing more detailed process diagrams of interactions [12] and implementing formal dynamic network models (for example, [13]) using time variables and canonical enzyme kinetics can enable the investigation of causal pathway dynamics. In this scenario, process diagrams or biochemical reaction networks can be translated into nonlinear differential equation models describing temporal changes of protein or gene expression concentrations. It is important to note that, inherently, biological pathways and their networks are nonlinear systems and as such their behavior cannot be described with a simple linear relationships between all components. For this reason dynamical systems modeling is an essential applied mathematical tool for understanding the behavior of complex systems over time [14]. Such models allow a wide range of questions to be addressed, such as how the long-term behavior of the system depends on initial activation conditions; how the coordinate regulation of a biosynthesis pathway influences the flux of intermediate metabolites; and to what extent and how cross talk between pathways retains specificity and avoids unwanted interference.

As examples, NF-κB signaling oscillation [15], lipid metabolism [16] and circadian oscillation [17] have all been effectively dynamically modeled using systems of ordinary differential equations that are relevant to macrophage function. Directly modeling pathway dynamics is complicated by the need for high-quality temporal data describing pathway activity and high-confidence parameter values, which can be directly obtained only from highly structured experimental data. Thus, quantitatively accurate pathway models are relatively rare. There are complementary modeling approaches that minimize these requirements, such as logic-based Boolean approaches (for a review of this approach, see [18,19]). However, these come with restrictions on the subtlety of behavior they can predict.

Several studies have modeled the macrophage at the level of the cell population. Its role in colonic inflammation [20], tumor growth and suppression (for example, [21]) and diabetes [22] have been explored *in silico*, and more speculative ideas from other fields, such as critical ordering in complex systems, have been explored using the macrophage as a test system [23].

Ultimately, *in silico* models could improve our understanding of how the molecular function of the macrophage dynamically generates the cellular function of macrophage activation, and this may lead to predictions of how best to intervene in order to modulate macrophage behavior (Figure 1).

**The manipulated macrophage**
So far, two principal approaches have been taken to manipulating macrophage function *in vivo*. The first has been to use genetic transduction of bone-marrow-derived cells. Macrophages and their progenitor stem cells can be transduced with recombinant viruses with high efficiency [24]. The second approach has been to use *ex vivo* stimulation by cytokines. Treatments such as these have a 'polarizing' effect on cellular phenotype, resulting in cells with functional characteristics appropriate to the clinical context and the treatment required [25]. In the future, microRNA mimics, small molecules and synthetic regulatory circuits could all conceivably be introduced as therapeutic devices based on the predictions of *in silico* models (as schematically outlined in Figure 1).

As well as being amenable to *ex vivo* manipulation, macrophages can also preferentially localize to the site of injury following *in vivo* administration. In experimental immune-mediated glomerulonephritis, transduced macrophages expressing IL-4, IL-10 or the IL-1 receptor antagonist, reduced inflammation and subsequent fibrosis [24,26]. Macrophages expressing such anti-inflammatory cytokines are known as *alternatively* activated and the injection of alternatively activated macrophages has been used to effectively treat experimental colitis [27]. Thus, across a range of disease model settings, exogenously administered manipulated macrophages can have beneficial therapeutic effects.

Other related types of cells may also have therapeutic potential, for example myeloid-derived suppressor cells (MDSCs), a heterogeneous group of cells that can downregulate T-cell-mediated immune responses. MDSC activity is important in supporting tumor growth [28] and suppressing tumor-associated immunity [29]. Thus, studies of MDSCs may provide insight into approaches to the downregulation of unwanted immune activity.

**Concluding remarks and future prospects**
Macrophages are highly flexible and adaptive cells of the immune system and are exquisitely sensitive to activation from a resting state by the direct sensing of pathogen and host danger signals in their microenvironment. The innate immune response and subsequent inflammatory reactions involve the macrophage initiating a complex gene expression program following activation. *In silico* modeling could enable studies of the molecular, genetic and proteomic function of the macrophage in relation to inflammation across multiple scales: molecular pathways in cell-autonomous immunity, intercellular communication pathways in tissue inflammation, and whole-organism response pathways in systemic disease.

One potential scenario is depicted in Figure 1. This illustrates how clinical research in inflammation and macrophage cell biology is moving towards the therapeutic clinical administration of modified macrophages (Figure 1a). In parallel (Figure 1b), laboratory studies are increasingly using high-throughput quantitative analyses of macrophage molecular function. Although still in its infancy, *in silico* simulation (Figure 1c) has a potentially key role in linking the laboratory and clinical areas by contributing a systems approach for precisely predicting the right mode of treatment to manipulate the macrophage phenotype for maximal therapeutic benefit. Although it might seem optimistic to imagine such a role for *in silico* simulation, we would argue that this is not too far in the future. Such progress, however, will require the refinement of techniques, the convergence of disciplines and the development of expertise in translation between disciplines if this vision is to become a reality.

Ultimately, *in silico* methods could supplant the financial and ethical cost associated with experimental bench work. However, near-term efforts will require the co-development of experimental and *in silico* models, enabling us to work with the underlying complexity to gain more insightful answers to the questions we pose than would be possible at the bench alone.

The opportunity is now emerging to harness *in silico* approaches to better understand macrophage biology in inflammatory diseases. Undoubtedly, this will accelerate the future exploitation of macrophage behavior in inflammatory disease.

**Abbreviations**
IL, interleukin; MDSC, myeloid-derived suppressor cells.

**Competing interests**
The authors declare that they have no competing interests.

**Author's contributions**
All authors contributed to the design and writing of this commentary.

**Acknowledgements**
This work was supported by the Wellcome Trust (WT066784) program grant and BBSRC to PG. The Centre for Systems Biology at Edinburgh is a Centre for Integrative Systems Biology supported by the BBSRC and EPSRC (BB/D019621/1).

**Figure 1**
Schematic diagram of how *in silico* macrophage modeling could be integrated with existing laboratory and therapeutic approaches. **(a)** Existing protocols can already reprogram patient macrophages e*x vivo* to treat inflammatory disease. Macrophages are taken from the patient; healthy macrophages are isolated; cytokines and/or chemokines are used to reprogram them; and they are then introduced back into the patient. **(b)** Laboratory investigations of high-throughput pathway-based analyses of multiple phenotypes are well established. The results of these analyses could be integrated with *in silico* simulations **(c)**, to predict effective treatments, such as small molecules or microRNAs. Testing these on macrophages isolated from patients (dashed arrow) in an iterative manner or 'systems loop' could be used to validate the *in silico* model. In this way, new, personalized phenotypic markers and macrophage reprogramming treatments (involving a single modification or a combination of modifications to cells) could be identified, and the therapeutic potential of the *ex vivo* cells will thereby be markedly enhanced.

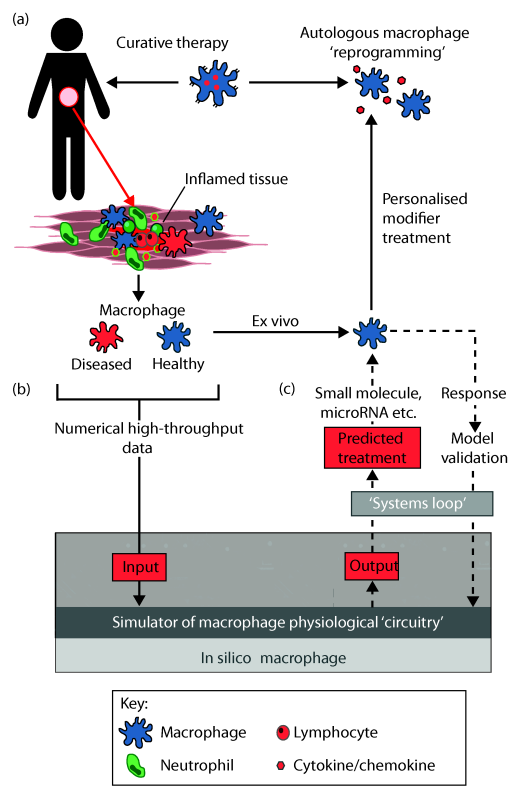

Figure 1